\documentclass[10pt, conference]{ieeeconf}     
\IEEEoverridecommandlockouts                           
\usepackage{graphicx}          
\usepackage{times,mathptmx}
\usepackage[T1]{fontenc}
\usepackage{color}
\usepackage{pstool}
\usepackage[latin1]{inputenc}
\usepackage{rotating}
\usepackage{amsmath}
\usepackage{amssymb}
\usepackage{graphicx}
\usepackage{macros}
\usepackage{booktabs}
\usepackage{url}
\usepackage{psfrag}
\usepackage{subfigure}
\usepackage{tikz}
\usepackage{pgfplots}

\renewcommand{\baselinestretch}{0.973}
\newcommand{\reals}{\mathbf R}

\title{\LARGE \bf   An ADMM Algorithm for Solving $\ell_1$ Regularized MPC}

\author{ Mariette Annergren$^*$, Anders Hansson$^{**}$, and Bo Wahlberg$^*$
\thanks{This work was partially supported by the Swedish Research Council and the Linnaeus Center ACCESS at KTH and the European Research Council under the advanced grant LEARN, contract 267381.}
\thanks{$^*$Automatic Control Lab and ACCESS, School of Electrical Engineering, KTH, SE-100 44 Stockholm, Sweden. (e-mail: \{mariette.annergren, bo.wahlberg\}@ee.kth.se)}
\thanks{$^{**}$Division of Automatic Control, Department of Electrical Engineering, Link\"{o}pings Universitet, SE-581 83 Link\"{o}ping, Sweden.
 (e-mail: anders.g.hansson@liu.se). This work was carried out when the
author was a Visiting Professor at University of California, Los Angeles.}}

\begin{document}

\newcommand{\figuretikz}[3]{\pgfplotsset{width=#1\columnwidth,height=#1\columnwidth,compat=newest,plot coordinates/math parser=false}\input{#1}}
\usetikzlibrary{shapes,arrows}
\tikzstyle{block} = [draw, fill=white, rectangle,
    minimum height=2.8em, minimum width=4.5em, node distance = 8em]
\tikzstyle{sum} = [draw, fill=white, circle, node distance=1cm, minimum size = 0.3cm]
\tikzstyle{input} = [coordinate]
\tikzstyle{output} = [coordinate]
\tikzstyle{pinstyle} = [pin edge={to-,thin,black}]
\tikzstyle{block2} = [draw,minimum width = 6em, minimum height=2.8em,text centered]

\maketitle
\thispagestyle{empty}
\pagestyle{empty}

\begin{abstract}          
We present an Alternating Direction Method of Multipliers (ADMM) algorithm for solving optimization problems with an $\ell_1$ regularized least-squares cost function subject to recursive equality constraints. The considered optimization problem has applications in control, for example in $\ell_1$ regularized MPC. The ADMM algorithm is easy to implement, converges fast to a solution of moderate accuracy, and enables separation of the optimization problem into sub-problems that may be solved in parallel. We show that the most costly step of the proposed ADMM algorithm is equivalent to solving an LQ regulator problem with an extra linear term in the cost function, a problem that can be solved efficiently using a Riccati recursion. We apply the ADMM algorithm to an example of $\ell_1$ regularized MPC. The numerical examples confirm fast convergence to moderate accuracy and a linear complexity in the MPC prediction horizon.       
\end{abstract}

\section{Introduction}
In this paper we consider optimization problems with an $\ell_1$ regularized least-squares cost function subject to recursive equality constraints. This has applications in control. The least squares part is standard in this context and penalizes deviations of the states from the set-point at the same time as keeping the control signal small. The $\ell_1$-norm regularization of the cost function promotes sparse solutions, i.e. a solution with many zero entries, \cite{boyd2004}. The cost function is known as LASSO, \cite{tibshirani1996}. LASSO is a well-known method in statistics and machine learning, and it has gained a lot of interest in other research communities as well, e.g. system identification, \cite{ohlsson2010}.

We propose to solve the optimization problem using an algorithm called Alternating Direction Method of Multipliers (ADMM). ADMM is a special case of  Douglas-Rachford splitting, \cite{eckstein1992}, and it is related to other optimization algorithms, e.g. method of multipliers and Bregman iterative algorithms for $\ell_1$ problems, \cite{gabay1976}, \cite{osher2005}. For an overview of ADMM, we refer the reader to \cite{boyd2011}.

The most costly step in the proposed ADMM algorithm is the projection of an iterate to a set describing a feasible solution. We will show that this projection is equivalent to solving a Linear Quadratic (LQ) regulator problem with an additional linear term in the cost function. This problem can be solved efficiently using a Riccati recursion just as in \cite{gla+jon84}. 

We will apply ADMM to the recently introduced $\ell_1$ regularized Model Predictive Control (MPC), \cite{maciejowski2012}. The $\ell_1$ regularized MPC has an $\ell_1$ regularized least-squares cost function. The motivation for $\ell_1$ regularized MPC is the reduced actuator activity obtained when using $\ell_1$-norm penalty on changes of the input signal \cite{maciejowski2012}. A detailed stability analysis of the closed loop system with $\ell_1$ regularized MPC, and results confirming sparse solutions are given in \cite{maciejowski2012}. In $\ell_1$ regularized MPC, an optimization problem such as the one we consider is solved at each sampling instant. Hence, the sampling time puts an upper bound on the time that the optimization is allowed to take, and therefore efficient algorithms are needed. It is believed that ADMM is a preferred algorithm for this application based on the result for LASSO, \cite{boyd2011}. We will see that this expectation is confirmed in numerical experiments.  

Also for traditional MPC an optimization problem has to be solved at each sampling instant, \cite{maciejowski2002}.  Because of this many different tailored optimization schemes have been developed to meet the real time requirements of MPC. Typically the optimization problem is a Quadratic Program (QP). There are mainly two different approaches that have been taken. One approach is to compute an explicit off-line solution to the QP which is stored in a look-up table, \cite{bem+mor+dua+pis02}. This facilitates very fast sampling, but is only feasible for small scale problems. The other approach is to compute the solution on-line as we propose, which is feasible also for medium- and large-scale problems. Among these approaches one can distinguish three different classes of methods: 1) Interior Point (IP) methods, \cite{rao+wri+raw97}, 2) Active-Set (AS) methods, \cite{gla+jon84} and 3) Fast Gradient (FG) methods, \cite{ric+jon+mor09}. 
Riccati recursions play an important role also in IP and AS methods for MPC, \cite{rao+wri+raw97,gla+jon84}, since they can be used for these methods to efficiently factorize the matrix involved in the linear system of equations for the search directions. So far they have not been used for FG methods. For IP methods the Riccati recursion has to be re-computed for each iterate of the method. For AS methods it has to be updated, i.e. parts of the old solution can be reused but has to be modified. For ADMM it is possible to use the same Riccati recursion for all iterates. Computing the Riccati recursion, i.e. factorizing the matrix for the search directions, is the most time-consuming task for all these methods. However, the convergence performance is not the same for the different methods, i.e. it takes a different amount of iterations to reach a solution of satisfactory accuracy. For IP methods the number of iterations is typically 20--50 to reach very high accuracy. For active set methods the number of iterations are typically higher, however by considering gradient projection methods on the dual problem speed can be gained, \cite{AxehillH:09}, and similar results as for IP methods can be obtained. For fast gradient methods it has in \cite{ric+jon+mor09} been shown how the number of iterates can be upper bounded to achieve a desired accuracy. Other recent relevant publications in relation to efficient methods for MPC include among others  \cite{fer+boc+die08,wan+boy10,pat+sop+sar11,wil+kna+nin12} and the references therein.

\section{Control problem}
We consider an open-loop control problem of finding an input sequence that minimizes a finite-horizon cost function, given a model and an initial state. The problem is formulated as follows
\begin{equation}\label{eq:opt1}
\begin{array}{ll}
\mbox{minimize} & \|x_H\|_{2,Q}^2+\sum_{i=1}^{H}\|y_{i-1}\|_{2}^2+\lambda \sum_{i=1}^{H}\|z_{i-1}\|_1,\\
\mbox{subject to} &  x_i = Ax_{i-1}+Bu_{i-1},\  i = 1,\ldots, H,\\
 & y_{i-1} = Cx_{i-1}+Du_{i-1},\  i = 1,\ldots, H,\\
 & z_{i-1}= Ex_{i-1}+Fu_{i-1},\  i = 1,\ldots, H,
\end{array}
\end{equation}
where $x_i\in\reals^n$ is the state vector, $u_i\in\reals^l$ is the input vector, $y_i\in\reals^m$ and $z_i\in\reals^p$ are auxiliary variables, and where $\|x\|^2_{2,A}=x^TAx$.  Formulation \eqref{eq:opt1} captures the optimization problems that may occur in $\ell_1$ regularized MPC. For example, we can replace the input vector with the change of the input by augmenting the state vector and modifying the system matrices accordingly, see \cite{maciejowski2002}.

\section{Alternating direction method of multipliers (ADMM)}
\label{sec:admm}
In this section, we provide a description of the key elements of ADMM. The description is a condensed version of the ones found in \cite{wahlberg2012} and \cite{boyd2011}. For a more rigorous overview, we refer the reader to \cite{boyd2011}.

\subsection{Optimization problem}
ADMM is a numerical algorithm for solving optimization problems such as
\begin{equation}\label{eq:admm_opt1}
\begin{array}{ll}
\mbox{minimize} & f(x),\\
\mbox{subject to} &  x \in {\mathcal C},
\end{array}
\end{equation}
for some vector variable $x\in\reals^n$, where $f(x)$ is a convex function and $\mathcal{C}$ is a convex set. An equivalent problem to \eqref{eq:admm_opt1} is
\begin{equation}\label{eq:admm_opt2}
\begin{array}{ll}
\mbox{minimize} & f(x) + I_\mathcal{C}(x_c),\\
\mbox{subject to} &  x = x_c,
\end{array}
\end{equation}
where $I_\mathcal{C}(x_c)$ is the indicator function of $\mathcal{C}$, \cite{boyd2004}.
\subsection{Augmented Lagrangian}
The augmented Lagrangian of optimization problem \eqref{eq:admm_opt2} is defined as
\begin{equation}\label{eq:admm_lagrangian}
L_\rho (x, x_c, x_d) = f(x) + I_\mathcal{C}(x_c) + (\rho/2)\|x-x_c+x_d\|_2^2,
\end{equation}
where $x_d$ is the dual variable corresponding to the equality constraint $x=x_c$ scaled by $1/\rho$, and $\rho>0$ is a tunable parameter. There is no simple way of finding the optimal $\rho$, however there are guidelines in \cite{boyd2011}.

\subsection{ADMM steps}
The ADMM algorithm consists of three main steps at each iteration $k$. The three steps are
\begin{align}
x^{k+1} &:= \argmin_x\{f(x) +(\rho/2)\|x-x_c^{k}+x_d^{k}\|_2^2\} \label{eq:admm_step1}\\
x_c^{k+1} &:= \Pi_{\mathcal{C}}(x^{k+1} + x_d^k) \label{eq:admm_step2},\\
x_d^{k+1} &:= x_d^k + (x^{k+1} - x_c^{k+1}) \label{eq:admm_step3},
\end{align}
where $\Pi_{\mathcal{C}}(x)$ denotes the Euclidean projection of a vector $x$ onto a set $\mathcal{C}$. In the first step \eqref{eq:admm_step1}, we minimize the augmented Lagrangian \eqref{eq:admm_lagrangian} with respect to $x$, keeping $x_c$ and $x_d$ fixed. In the second step \eqref{eq:admm_step2}, we minimize the augmented Lagrangian \eqref{eq:admm_lagrangian} with respect to $x_c$, keeping $x$ and $x_d$ fixed. In the third and last step \eqref{eq:admm_step3}, we update the scaled dual variable $x_d$. We then repeat all three steps until convergence. For more details and a complete convergence analysis, we refer the reader to \cite{boyd2011}.

\subsection{Stopping criteria}
\label{sec:stop}
The ADMM algorithm is iterated until some stopping criteria are fulfilled. We use criteria based on the primal and dual residuals of the optimization problem. The primal and dual residuals of \eqref{eq:admm_opt2} are 
\[
e_p^k = (x^k-x_c^k), \quad
e_d^k = -\rho (x_c^k-x_c^{k-1}).
\]
We terminate the algorithm when 
\begin{equation}\label{eq:admm_stop}
\begin{array}{rl}
\|e_p^k\|_2 &\!\!\!\!\!\!\!\leq \sqrt{n} \epsilon^\mathrm{abs} + \epsilon^\mathrm{rel}\max\{\|x^k\|_2, \|x_c^k\|_2\},\\
\|e_d^k\|_2 &\!\!\!\!\!\!\!\leq \sqrt{n} \epsilon^\mathrm{abs} + \epsilon^\mathrm{rel}\rho \|x_d^k\|_2,
\end{array}
\end{equation}
where $\epsilon^\mathrm{abs} > 0$ and $\epsilon^\mathrm{rel} > 0$ are absolute
and relative tolerances, respectively. For more details, see \cite{boyd2011}.

\subsection{Over-relaxation}
We may use over-relaxation to improve convergence of the ADMM algorithm, \cite{eckstein1992}, \cite{eckstein1994}, \cite{eckstein1998}. When using over-relaxation we modify the update $x^{k+1}$ with
\[
\hat{x}^{k+1} = \alpha  x^{k+1} + (1-\alpha) x_c^k,
\]
where $1.5\leq\alpha\leq1.8$, in the second and the third ADMM steps, \eqref{eq:admm_step2} and \eqref{eq:admm_step3}. For more details, see \cite{boyd2011}.
\section{Problem formulation and method}
\label{sec:problem}
In this section, we describe how the considered optimization problem in \eqref{eq:opt1} can be solved using ADMM.

\subsection{ADMM formulation}
The optimization problem in \eqref{eq:opt1} is on the same form as the optimization problem in \eqref{eq:admm_opt1}. The vector variables are
\begin{equation}\nonumber
\begin{array}{lll}
&x=(x_0,\ldots,x_H),&y=(y_0,\ldots,y_{H-1}),\\
& u=(u_0,\ldots,u_{H-1}),&z=(z_0,\ldots,z_{H-1}),\\
\end{array}
\end{equation}
the objective function is
\begin{align}\nonumber
f(x,y,u,z)=&\|x_H\|_{2,Q}^2+\sum_{i=1}^{H}\|y_{i-1}\|_{2}^2+\lambda \sum_{i=1}^{H}\|z_{i-1}\|_1,
\end{align}
and the constraint set is given by
\begin{equation}\nonumber
\begin{array}{ll}
\mathcal{C} = \{(x,y,u,z)|\!\!\!\!\!\!\!&x_i = Ax_{i-1} +Bu_{i-1} ,\  i = 1,\ldots, H\\
 & y_{i-1} = Cx_{i-1}+Du_{i-1},\  i = 1,\ldots, H,\\
 & z_{i-1}= Ex_{i-1}+Fu_{i-1},\  i = 1,\ldots, H\}.
\end{array}
\end{equation}
Thus, the ADMM formulation of optimization problem in \eqref{eq:opt1} is
\begin{equation}\label{eq:admm_opt3}
\begin{array}{ll}
\mbox{minimize} & f(x,y,u,z)+ I_\mathcal{C}(x_c,y_c,u_c,z_c),\\
\mbox{subject to} &  x = x_c,\ y = y_c,\ u = u_c,\ z = z_c.
\end{array}
\end{equation}

\subsection{Step 1 of ADMM}
The first ADMM step, \eqref{eq:admm_step1}, is almost the same as the one for $\ell_1$ mean filtering in \cite{wahlberg2012}. We solve $4H+1$ separate minimization problems because $f(x,y,u,z)$ is separable in its arguments. For the vector variables $x$, $y$ and $u$ the minimization problems have a quadratic cost function and no constraints. The solutions are
\begin{align}\nonumber
x^{k+1}_i&= x_{c,i}^{k}-x_{d,i}^{k},\ i=0,\ldots,H-1,\\ \nonumber
x^{k+1}_H&= (2Q+\rho I_{n})^{-1}\rho(x_{c,H}^{k}-x_{d,H}^{k}),\\\nonumber
y^{k+1}_i&= (2+\rho)^{-1}\rho(y_{c,i}^{k}-y_{d,i}^{k}),\ i=0,\ldots,H-1,\\\nonumber
u^{k+1}_i&= u_{c,i}^{k}-u_{d,i}^{k},\ i=0,\ldots,H-1,
\end{align}
where $I_a$ denotes the identity matrix in $\reals^{a\times a}$. For the vector variable $z$, the minimization problems are
\begin{equation}\label{eq:admm_optZ}
z^{k+1}_i = \argmin_{z_i}\{\lambda\|z_i\|_1+(\rho/2)\|z_i-z_{c,i}^{k}+z_{d,i}^{k}\|_2^2\},
\end{equation}
with component-wise solutions $(z_i^{k+1})_j =\mathcal{S}_{\lambda/\rho}((z_{c,i}^{k}-z_{d,i}^{k})_j),$ for $i=0,\ldots,H-1,$ and $j=1,\ldots,p,$ where $\mathcal{S}_{\lambda/\rho}$ denotes the soft thresholder operator, see \cite{boyd2011}.

\subsection{Step 2 of ADMM}
\label{sec:proj}
The second step of ADMM, \eqref{eq:admm_step2}, consists of a projection of the vector
\[(x_p^{k},y_p^{k},u_p^{k},z_p^{k})=(x^{k+1} + x_d^k, y^{k+1} + y_d^k,u^{k+1} + u_d^k, z^{k+1} + z_d^k)\]
onto the constraint set $\mathcal{C}$, i.e.
\[
(x_c^{k+1}, y_c^{k+1},u_c^{k+1}, z_c^{k+1}) = \Pi_\mathcal{C}((x_p^{k},y_p^{k},u_p^{k},z_p^{k})).
\]
The projection can be formulated as the optimization problem 
\begin{equation}\label{eq:admm_proj}
\begin{array}{ll}
\mbox{minimize} & \|(x_c^{k+1}, y_c^{k+1},u_c^{k+1}, z_c^{k+1})-(x_p^{k},y_p^{k},u_p^{k},z_p^{k})\|_2^2,\\
\mbox{subject to} &  (x_c^{k+1}, y_c^{k+1},u_c^{k+1}, z_c^{k+1})\in\mathcal{C}.
\end{array}
\end{equation}
To simplify notation in the rest of this section, we will drop the use of super script $k$ and $k+1$. An equivalent optimization problem to the one in \eqref{eq:admm_proj} is
\begin{equation}\label{eq:admm_proj2}
\begin{array}{ll}
\mbox{minimize} & v^T\mathcal{Q}v+q^Tv\\
\mbox{subject to} &   \mathcal{F}v=g,
\end{array}
\end{equation}
where 
\begin{align*}\nonumber
v&= (x_{c,0},u_{c,0},\ldots,x_{c,H-1},u_{c,H-1},x_{c,H}),\\
g& = (x_{c,0},0,\ldots,0),\\
q&= (r_0,s_0,\ldots,r_{H-1},s_{H-1},0),\\
r_i & = -2(x_{p,i}^T + z_{p,i}^TE + y_{p,i}^TC),\\
s_i & = -2(u_{p,i}^T + z_{p,i}^TF+y^T_{p,i}D),\\
\mathcal{Q}&=
 \left[
\begin{array}{ll}
T& 0\\
0 & Q\\
\end{array}
\right],\quad
T=I_{H}\otimes
 \left[
\begin{array}{ll}
P& S\\
S^T & R\\
\end{array}
\right],\\
P & = I_{n} + C^TC + E^TE,\\
R & = I_{l} + D^TD+F^TF,\\
S & = C^TD+E^TF,\\
\mathcal{F}&=
 \left[
\begin{array}{llllll}
I_{n} & 0&0&0&\ldots&0 \\
-A & -B&I_{n}&0&\ldots&0 \\
\vdots & \ddots&\ddots&\ddots&\ddots&\vdots \\
0 & \ldots&\dots&-A&-B&I_{n}\\
\end{array}
\right].
\end{align*}
The symbol $\otimes$ denotes the Kronecker product. The optimization problem in \eqref{eq:admm_proj2} is an equality constrained minimization problem. As such, its solution is equivalent to the solution of its Karush-Kuhn-Tucker (KKT) conditions, \cite{boyd2004}. The KKT conditions of the optimization problem in \eqref{eq:admm_proj2} are
\begin{equation}\label{eq:admm_kkt}
\left[
\begin{array}{ll}
2\mathcal{Q} & \mathcal{F}^T\\
\mathcal{F}&0 \\
\end{array}
\right]w-\left[
\begin{array}{l}
-q\\
g\\
\end{array}
\right]=0,
\end{equation}
with $w = (v, v_d)$, where $v_d$ is the Lagrange multiplier corresponding to the equality constraint $\mathcal{F}v=g$. The KKT conditions in \eqref{eq:admm_kkt} are a system of linear equations and can be efficiently solved using a Riccati recursion as described in the Appendix. The solution to the optimization problem in \eqref{eq:admm_proj} is obtained by extracting $x_c$ and $u_c$ from $v$, and calculating $y_c$ and $z_c$ from the equations defining the constraint set $\mathcal{C}$.

\subsection{Step 3 of ADMM}
In the third ADMM step in \eqref{eq:admm_step3}, we update the scaled dual variables, i.e.
\begin{align*}\nonumber
x_{d,i}^{k+1} &= x_{d,i}^k + (x_i^{k+1}-x_{c,i}^{k+1}), \quad i = 0,\ldots,H,\\
y_{d,i}^{k+1}& = y_{d,i}^k + (y_i^{k+1}-\tilde{y}_{c,i}^{k+1}), \quad i = 0,\ldots,H-1,\\
u_{d,i}^{k+1} &= u_{d,i}^k + (u_i^{k+1}-u_{c,i}^{k+1}), \quad i = 0,\ldots,H-1,\\
z_{d,i}^{k+1}& = z_{d,i}^k + (z_i^{k+1}-z_{c,i}^{k+1}), \quad i = 0,\ldots,H-1.
\end{align*}

\section{Example}
In this section, we describe how the ADMM algorithm, proposed in \ref{sec:problem}, can be used to solve an $\ell_1$ regularized MPC problem without inequality constraints.

\subsection{Model}
\label{sec:admm}
We consider a linear and discrete model of the plant. The model is given by
\begin{eqnarray}\label{eq:mpc_model}
\begin{array}{rl}
x(t+1)&\!\!\!\!\!\!\!= Ax(t)+Bu(t),\\
y(t)&\!\!\!\!\!\!\!= Cx(t),
\end{array}
\end{eqnarray}
where $x(t)\in\reals^n$ is the state vector, $u(t)\in\reals^l$ is the input vector and  $y(t)\in\reals^m$ is the output vector.

\subsection{Cost function}
The control objective is to drive the output vector to zero, namely the regulator problem \cite{maciejowski2002}, while using a piece-wise constant input signal. Such a control objective can be described by the cost function
\begin{equation}\label{eq:mpc_cost}
\begin{array}{ll}
V(t) =&\!\!\!\!\!\!\! \|\hat{x}(t+H_p|t)\|_{2,\bar{Q}}^2+\sum_{i=1}^{H_p}\|\hat{y}(t+i-1|t)\|_{2,Q}^2+\\
&\!\!\!\!\!\!\!\lambda \sum_{i=1}^{H_u}\!\!\|\Delta\hat{u}(t+i-1|t)\|_1.
\end{array}
\end{equation} 
The cost function penalizes the terminal state, output deviation from zero and non-constant input signals. In \eqref{eq:mpc_cost}, $\hat{x}(t+i|t)$ and $\hat{y}(t+i|t)$ are the predicted state and output vectors, respectively, at time $t+i$ given measurements up to time $t$ and the  model in \eqref{eq:mpc_model}. Moreover, 
\[\Delta\hat{u}(t+i|t)=\hat{u}(t+i|t)-\hat{u}(t+i-1|t),\]
where $\hat{u}(t+i|t)$ is the predicted input vector given measurements up to time $t$ and the  model in \eqref{eq:mpc_model}. The prediction and control horizons are denoted $H_p$ and $H_u$ respectively, and we assume that $\Delta\hat{u}(t+i|t)=0$ for all $i\geq H_u$. The matrices $Q,\bar{Q}\in\reals^{m\times m}$ and the scalar $\lambda$ are weights. We require that $Q$ and $\bar{Q}$ are positive semidefinite, and that $\lambda$ is non-negative. 
\subsection{Optimization problem}
The control objective can be achieved by minimizing the cost function in \eqref{eq:mpc_cost} given the model in \eqref{eq:mpc_model} in each time step $t$, in accordance with the receding horizon idea, \cite{maciejowski2002}. We can formulate the optimization problem as
\begin{equation}\label{eq:mpc_opt}
\begin{array}{ll}
\mbox{minimize} &  V(t),\\
\mbox{subject to} &  \hat{x}(t+i|t) = A\hat{ x}(t+i-1|t)+B\hat{ u}(t+i-1|t),\\
& i = 1,\ldots, H_p,\\
& \hat{x}(t|t) = x(t),\\
&\hat{y}(t+i-1|t) = C\hat{x}(t+i-1|t),\  i = 1,\ldots, H_p.
\end{array}
\end{equation}
The optimization problem in \eqref{eq:mpc_opt} is similar to standard formulations as the one found in \cite{maciejowski2002}. The significant difference is the use of the $\ell_1$-norm of  $\Delta\hat{u}(t+i|t)$ instead of the $\ell_2$-norm in the cost function in \eqref{eq:mpc_cost}. The former typically promotes sparse $\Delta\hat{u}(t+i|t)$ for $i=0,\ldots, H_u-1$, while the latter promotes small but non-zero elements of $\Delta\hat{u}(t+i|t)$ for $i=0,\ldots, H_u-1$, \cite{boyd2004}. To simplify notation in the rest of the paper, we denote $\hat{x}(t+i|t)$ as $x_i$, $\hat{y}(t+i|t)$ as $y_i$, and so forth.

\subsection{Receding horizon}
The optimization problem in \eqref{eq:mpc_opt} is solved with respect to the input vector $u_i$ for $i=0,\ldots, H_u-1$. The input vector at the first time step, $u_0$, is applied to the plant. The state vector is updated according to measurements and, if necessary, an observer. The optimization problem in \eqref{eq:mpc_opt} is updated and solved again. The described procedure is repeated until some final time step. Note that closed loop stability cannot be guaranteed for all values of $\lambda$, see \cite{maciejowski2012}. Typically, a terminal cost penalty is used to obtain closed loop stability, if possible.

\subsection{MPC formulation}
We consider the optimization problem in \eqref{eq:mpc_opt}. We set the predicted output vector to be the predicted state vector, and the prediction horizon equal to the control horizon, that is,
\begin{equation}\label{eq:mpc_opt1}
\begin{array}{ll}
\mbox{minimize}& \|x_H\|_{2,\bar{Q}}^2+\sum_{i=1}^{H}\|x_{i-1}\|_{2,Q}^2+\lambda \sum_{i=1}^{H}\|\Delta u_{i-1}\|_1,\\
\mbox{subject to} &  x_i = Ax_{i-1}+Bu_{i-1},\  i = 1,\ldots, H.
\end{array}
\end{equation}
The optimization problem in \eqref{eq:mpc_opt1} can be reformulated by replacing $u_i$ with $\Delta u_i$ in a similar way as in \cite{maciejowski2002}. We introduce three new vector variables $\tilde{x}_i\in \reals^{n+l}$, $\tilde{y}_i\in\reals^{n}$ and $z_i\in\reals^l$ in the following way
\begin{equation}\label{eq:mpc_opt2}
\begin{array}{ll}
\mbox{minimize} & \|\tilde{x}_H\|_{2,\tilde{Q}}^2+\sum_{i=1}^{H}\|\tilde{y}_{i-1}\|_{2}^2+\lambda \sum_{i=1}^{H}\|z_{i-1}\|_1,\\
\mbox{subject to} &  \tilde{x}_i = \tilde{A}\tilde{x}_{i-1}+\tilde{B}\Delta u_{i-1},\  i = 1,\ldots, H,\\
 & \tilde{y}_{i-1} = \tilde{C}\tilde{x}_{i-1}+D\Delta u_{i-1},\  i = 1,\ldots, H,\\
 & z_{i-1}= E\tilde{x}_{i-1}+F\Delta u_{i-1},\  i = 1,\ldots, H.
\end{array}
\end{equation}
Here, $\tilde{x}_i$ is the state vector augmented with the input vector at the previous time step, i.e. $\tilde{x}_i=(x_i,u_{i-1})$. The matrices $\tilde{A}$, $\tilde{B}$ and $\tilde{C}$ are given by
\[
\tilde{A}= \left[
\begin{array}{ll}
A& B\\
0 & I_l\\
\end{array}
\right],\quad
\tilde{B}= \left[
\begin{array}{l}
B\\
I_l\\
\end{array}
\right],\quad
\tilde{C}= \left[
\begin{array}{ll}
c& 0\\
\end{array}
\right],
\]
where $c$ is chosen such that $Q = c^Tc$, and the matrices $D$, $E$, $F$ and $\tilde{Q}$ are given by
\[
D=0,\quad
E=0,\quad
F= I_l,\quad
\tilde{Q}= \left[
\begin{array}{ll}
\bar{Q}& 0\\
0 & 0\\
\end{array}
\right].
\]
The optimization problem in \eqref{eq:mpc_opt2} is of the same form as the one in \eqref{eq:opt1}, with $\Delta u_i$ acting as the input. Therefore, we can solve \eqref{eq:mpc_opt2} efficiently using ADMM and a Riccati recursion as described in  Section~\ref{sec:problem}. Note that we in this particular case can pre-calculate $F_i$, $H_i$, $G_i$, and $S_i$ in the Riccati recursion, before we start the MPC iterations.

\section{Numerical examples}
\label{sec:example}
In this section, we apply the ADMM algorithm on an $\ell_1$ regularized MPC problem. All the examples are performed with $\rho=1$. To improve convergence we use over-relaxation with $\alpha=1.8$ and we warm-start each ADMM iteration with the variable values obtained in the previous MPC iteration. We use stopping criteria \eqref{eq:admm_stop} with $\epsilon^{abs}=10^{-5}$ and $\epsilon^{rel}=10^{-4}$.

\subsection{Example 1: Quadruple water tank process}
\label{sec:ex2}

\subsubsection{Plant}
\begin{figure}[ht]
\begin{center}
\includegraphics[trim = 0mm 0mm 0mm 0mm, clip,width = 0.78\columnwidth]{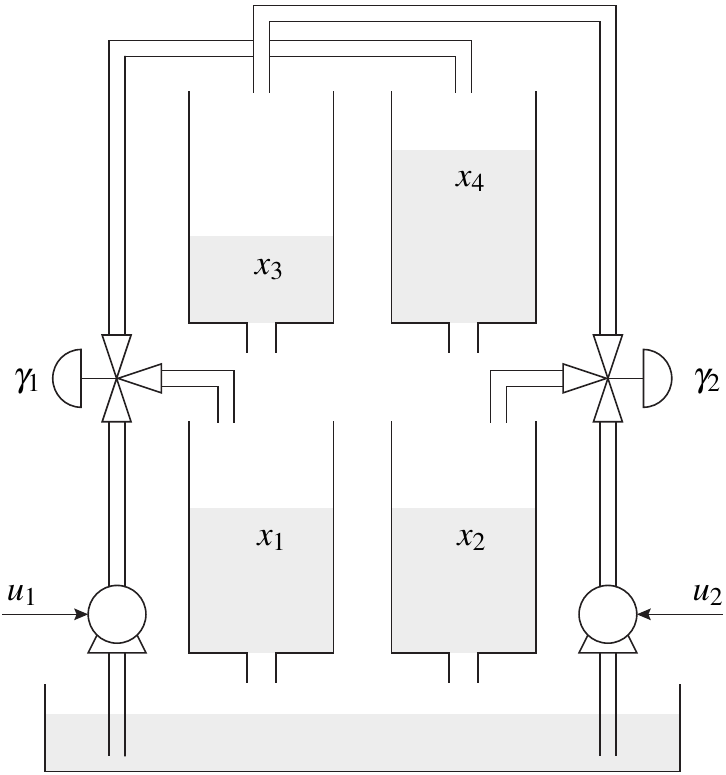}
\caption{Quadruple water tank process. }\label{fig:process}\end{center}
\end{figure}

The plant is the quadruple water tank process presented in \cite{hansson1999}. The process is shown in Figure \ref{fig:process}, where $x=(x_1,x_2,x_3,x_4)$ are the water levels, $u=(u_1,u_2)$ are the pump voltages, and $\gamma_1=\gamma_2=0.625$ are the parameters associated with the valves. The area of the cross-sections of the outlets of each tank are $(a_1,a_2,a_3,a_4)=$(0.17,0.15,0.11,0.08) cm$^2$, the area of the cross-sections of each tank are $A=15.5$ cm$^2$ and the parameters associated with the pumps are $k_1=k_2=4.14$ cm$^3$/(sV). 

\subsubsection{Model}
We obtain a linear model of the process by linearizing the nonlinear plant description given in \cite{hansson1999} around its equilibrium points. The linearized model is 
\begin{align*}
\frac{d\bar{x}}{dt}&=\left[\begin{matrix}
\frac{-1}{\tau_1}&0&\frac{1}{\tau_3}&0\\
0&\frac{-1}{\tau_2}&0&\frac{1}{\tau_4}\\
0&0&\frac{-1}{\tau_3}&0\\
0&0&0&\frac{-1}{\tau_4}
\end{matrix}\right]\bar{x}_t+\left[\begin{matrix}
\frac{\gamma_1k_1}{A}&0\\
0&\frac{\gamma_2k_2}{A}\\
0&\frac{(1-\gamma_2)k_2}{A}\\
\frac{(1-\gamma_1)k_1}{A}&0
\end{matrix}\right]\bar{u}_t,\\
\label{eq:Lin2}
y &= \left[\begin{matrix}
1&0&0&0\\
0&1&0&0
\end{matrix}\right]\bar{x}_t,
\end{align*}
where $\bar{x}=x-x^0$, $\bar{u}=u-u^0$ and $\tau_i=\frac{A}{a_i}\sqrt{\frac{2x_i^0}{g}}$. The equilibrium points of the plant are $x^0=(15,15,3,12)$ cm and $u^0=(7.8,5.25)$ V. The linear model is discretized assuming zero-order hold sampling at a sampling rate of 1 Hz.

\subsubsection{Simulation}
We set $H=5$, $Q=I_2$, and $\tilde{Q}=0$. The plant is initialized with $x(0)=(16,16,4,13)$ cm and $u(0)=u^0$. A Kalman filter is used to estimate the complete state vector during simulation. The MPC iterates for 10 time steps. We perform the same MPC simulation for $\lambda$ equal to 0.05, 0.1, 2 and 5. The applied input sequences are shown in Figure \ref{fig:ex2_inputs} and the output sequences are shown in Figure \ref{fig:ex2_outputs}. We see that the applied input signal varies over time for low values of $\lambda$. As $\lambda$ gets larger, the input signal becomes piece-wise constant, and eventually completely constant. We also see that a more restrictive control strategy, i.e. a high value of $\lambda$, gives worse control performance in terms of response time and static error.
\begin{figure}[th!]
\centering
\pgfplotsset{width=ex1_inputs.tikz\columnwidth,height=ex1_inputs.tikz\columnwidth,compat=newest,plot coordinates/math parser=false}
%
%
\begin{tikzpicture}

\begin{axis}[%
name=plot1,
scale only axis,
width=6cm,
height=4cm,
xmin=0, xmax=10,
ymin=5, ymax=8,
ylabel={$u_1(t)$},
axis on top]
\addplot[const plot,color=black,dashed,line width=1.0pt] plot coordinates{ (0,7.8) (1,5.03884) (2,5.66065) (3,7.08078) (4,7.34924) (5,7.53816) (6,7.63033) (7,7.64489) (8,7.66742) (9,7.6853) (10,7.69525) };
\label{dashed}
\addplot[const plot,color=black,dotted,line width=1.0pt] plot coordinates{ (0,7.8) (1,6.20416) (2,6.20415) (3,6.20427) (4,6.76917) (5,7.51447) (6,7.56026) (7,7.59953) (8,7.63802) (9,7.66642) (10,7.68334) };
\label{dotted}
\addplot[const plot,color=black,dash pattern=on 1pt off 3pt on 3pt off 3pt,line width=1.0pt] plot coordinates{ (0,7.8) (1,7.50725) (2,7.18212) (3,7.16821) (4,7.16821) (5,7.16826) (6,7.16831) (7,7.16836) (8,7.16834) (9,7.1684) (10,7.16845) };
\label{dashed2}
\addplot[const plot,color=black,solid,line width=1.0pt] plot coordinates{ (0,7.8) (1,7.80018) (2,7.80014) (3,7.80004) (4,7.79997) (5,7.79992) (6,7.79992) (7,7.79994) (8,7.8) (9,7.79996) (10,7.80002) };
\label{solid}
\end{axis}

\begin{axis}[%
at=(plot1.below south west), anchor=above north west,
scale only axis,
width=6cm,
height=4cm,
xmin=0, xmax=10,
ymin=2, ymax=6,
xlabel={$t$},
ylabel={$u_2(t)$},
axis on top]
\addplot[const plot,color=black,dashed,line width=1.0pt] plot coordinates{ (0,5.25) (1,2.11186) (2,2.5721) (3,4.92351) (4,5.12923) (5,5.14374) (6,5.16266) (7,5.17851) (8,5.19271) (9,5.20202) (10,5.20552) };

\addplot[const plot,color=black,dotted,line width=1.0pt] plot coordinates{ (0,5.25) (1,3.42892) (2,3.42892) (3,3.42864) (4,4.85486) (5,5.05503) (6,5.10327) (7,5.13827) (8,5.16497) (9,5.18256) (10,5.19157) };

\addplot[const plot,color=black,dash pattern=on 1pt off 3pt on 3pt off 3pt,line width=1.0pt] plot coordinates{ (0,5.25) (1,4.84676) (2,4.53176) (3,4.53184) (4,4.53188) (5,4.53194) (6,4.53201) (7,4.53207) (8,4.53205) (9,4.5321) (10,4.53216) };

\addplot[const plot,color=black,solid,line width=1.0pt] plot coordinates{ (0,5.25) (1,5.25027) (2,5.25021) (3,5.25021) (4,5.25023) (5,5.25032) (6,5.25044) (7,5.25056) (8,5.25069) (9,5.25061) (10,5.25072) };

\end{axis}
\end{tikzpicture}
\caption{Applied input sequences. The applied input sequence denoted (\ref{dashed}), (\ref{dotted}), (\ref{dashed2}) and (\ref{solid}) corresponds to $\lambda$ equal to 0.05, 0.1, 2 and 5 respectively.}\label{fig:ex2_inputs}
\end{figure}
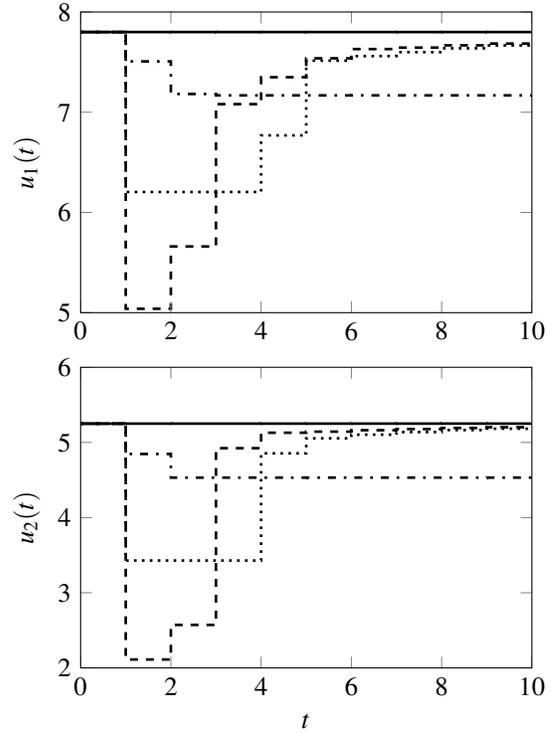
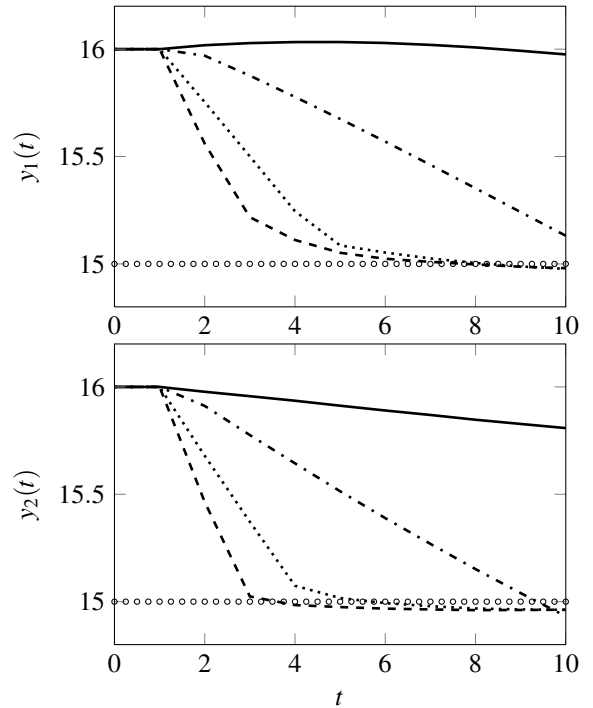
\begin{figure}[th!]
\centering
\pgfplotsset{width=ex1_outputs.tikz\columnwidth,height=ex1_outputs.tikz\columnwidth,compat=newest,plot coordinates/math parser=false}
%
%
\begin{tikzpicture}

\begin{axis}[%
name=plot1,
scale only axis,
width=6cm,
height=4cm,
xmin=0, xmax=10,
ymin=14.8, ymax=16.2,
ylabel={$y_1(t)$},
axis on top]
\addplot [
color=black,
dashed,
line width=1.0pt
]
coordinates{
 (0,16)(1,16)(2,15.5599)(3,15.2176)(4,15.1114)(5,15.0518)(6,15.0239)(7,15.0096)(8,14.9974)(9,14.9869)(10,14.9798) 
};

\addplot [
color=black,
dotted,
line width=1.0pt
]
coordinates{
 (0,16)(1,16)(2,15.7534)(3,15.5005)(4,15.2455)(5,15.0862)(6,15.0529)(7,15.0255)(8,15.0044)(9,14.988)(10,14.9769) 
};

\addplot [
color=black,
dash pattern=on 1pt off 3pt on 3pt off 3pt,
line width=1.0pt
]
coordinates{
 (0,16)(1,16)(2,15.9695)(3,15.8766)(4,15.7775)(5,15.675)(6,15.5697)(7,15.4614)(8,15.3522)(9,15.2412)(10,15.1311) 
};

\addplot [
color=black,
solid,
line width=1.0pt
]
coordinates{
 (0,16)(1,16)(2,16.0184)(3,16.0282)(4,16.033)(5,16.0331)(6,16.0288)(7,16.0199)(8,16.0081)(9,15.9924)(10,15.9755) 
};

\addplot [
color=black,
mark size=1.0pt,
only marks,
mark=o,
mark options={solid}
]
coordinates{
 (0,15)(0.25,15)(0.5,15)(0.75,15)(1,15)(1.25,15)(1.5,15)(1.75,15)(2,15)(2.25,15)(2.5,15)(2.75,15)(3,15)(3.25,15)(3.5,15)(3.75,15)(4,15)(4.25,15)(4.5,15)(4.75,15)(5,15)(5.25,15)(5.5,15)(5.75,15)(6,15)(6.25,15)(6.5,15)(6.75,15)(7,15)(7.25,15)(7.5,15)(7.75,15)(8,15)(8.25,15)(8.5,15)(8.75,15)(9,15)(9.25,15)(9.5,15)(9.75,15)(10,15) 
};

\end{axis}

\begin{axis}[%
at=(plot1.below south west), anchor=above north west,
scale only axis,
width=6cm,
height=4cm,
xmin=0, xmax=10,
ymin=14.8, ymax=16.2,
xlabel={$t$},
ylabel={$y_2(t)$},
axis on top]
\addplot [
color=black,
dashed,
line width=1.0pt
]
coordinates{
 (0,16)(1,16)(2,15.4639)(3,15.0244)(4,14.9841)(5,14.9748)(6,14.9676)(7,14.9638)(8,14.9605)(9,14.9612)(10,14.9625) 
};

\addplot [
color=black,
dotted,
line width=1.0pt
]
coordinates{
 (0,16)(1,16)(2,15.6794)(3,15.3717)(4,15.0742)(5,15.0169)(6,14.9933)(7,14.9788)(8,14.9682)(9,14.9639)(10,14.9618) 
};

\addplot [
color=black,
dash pattern=on 1pt off 3pt on 3pt off 3pt,
line width=1.0pt
]
coordinates{
 (0,16)(1,16)(2,15.9115)(3,15.7759)(4,15.6441)(5,15.5141)(6,15.3891)(7,15.2692)(8,15.1516)(9,15.04)(10,14.932) 
};

\addplot [
color=black,
solid,
line width=1.0pt
]
coordinates{
 (0,16)(1,16)(2,15.9774)(3,15.9568)(4,15.9357)(5,15.9125)(6,15.8902)(7,15.8694)(8,15.8472)(9,15.8276)(10,15.8082) 
};

\addplot [
color=black,
mark size=1.0pt,
only marks,
mark=o,
mark options={solid}
]
coordinates{
 (0,15)(0.25,15)(0.5,15)(0.75,15)(1,15)(1.25,15)(1.5,15)(1.75,15)(2,15)(2.25,15)(2.5,15)(2.75,15)(3,15)(3.25,15)(3.5,15)(3.75,15)(4,15)(4.25,15)(4.5,15)(4.75,15)(5,15)(5.25,15)(5.5,15)(5.75,15)(6,15)(6.25,15)(6.5,15)(6.75,15)(7,15)(7.25,15)(7.5,15)(7.75,15)(8,15)(8.25,15)(8.5,15)(8.75,15)(9,15)(9.25,15)(9.5,15)(9.75,15)(10,15) 
};
\label{circle}
\end{axis}
\end{tikzpicture}
\caption{Measured output sequences. The measured output sequence denoted (\ref{dashed}), (\ref{dotted}), (\ref{dashed2}) and (\ref{solid}) corresponds to $\lambda$ equal to 0.05, 0.1, 2 and 5 respectively. The sequence (\ref{circle} \ref{circle} \ref{circle}) corresponds to the equilibrium point of the water levels.}\label{fig:ex2_outputs}
\end{figure}

\subsection{Example 2: Number of iterations in ADMM}
\begin{figure}[th!]
\centering
\pgfplotsset{width=ex6_iter.tikz\columnwidth,height=ex6_iter.tikz\columnwidth,compat=newest,plot coordinates/math parser=false}
%
%
\begin{tikzpicture}

\begin{axis}[%
scale only axis,
width=6cm,
height=4cm,
xmin=0, xmax=10,
ymin=0, ymax=500,
xlabel={$t$},
ylabel={iterations in ADMM},
axis on top]
\addplot [
color=black,
dashed,
line width=1.0pt
]
coordinates{
 (1,465)(2,181)(3,237)(4,160)(5,142)(6,63)(7,67)(8,67)(9,67)(10,67) 
};

\addplot [
color=black,
dotted,
line width=1.0pt
]
coordinates{
 (1,263)(2,205)(3,99)(4,214)(5,129)(6,85)(7,82)(8,78)(9,78)(10,74) 
};

\addplot [
color=black,
dash pattern=on 1pt off 3pt on 3pt off 3pt,
line width=1.0pt
]
coordinates{
 (1,125)(2,96)(3,82)(4,57)(5,59)(6,59)(7,59)(8,60)(9,59)(10,59) 
};

\addplot [
color=black,
solid,
line width=1.0pt
]
coordinates{
 (1,77)(2,65)(3,58)(4,53)(5,46)(6,42)(7,42)(8,42)(9,43)(10,42) 
};

\end{axis}
\end{tikzpicture}
\caption{Number of iterations in ADMM required for fulfilling the stopping criteria. The iteration sequence denoted (\ref{dashed}), (\ref{dotted}), (\ref{dashed2}) and (\ref{solid}) corresponds to $\lambda$ equal to 0.05, 0.1, 2 and 5 respectively. The number of iterations required drops when no rapid changes in plant inputs or states occur.}\label{fig:ex6_iter}
\end{figure}
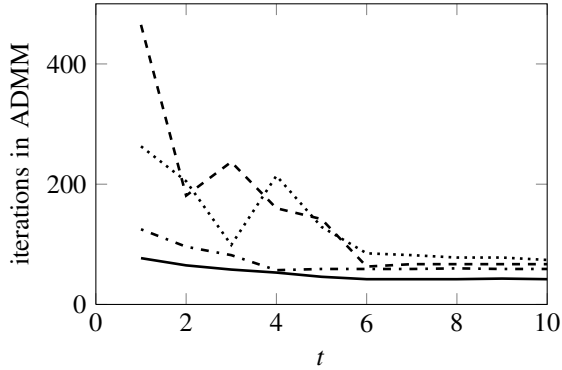

Figure \ref{fig:ex6_iter} shows the number of iterations required in ADMM for fulfilling the stopping criteria in Example 1. We also investigated the number of iterations required without warm-starting the algorithm. The conclusion is that a warm-start improves the convergence of ADMM when the plant inputs are close to constant and no rapid changes in the plant states occur. When this is not the case, we get similar performance with and without warm-start. It is natural that the benefit of warm-starting is greater the less the states move. 

\subsection{Example 3: Convergence of ADMM}
\label{sec:ex3}
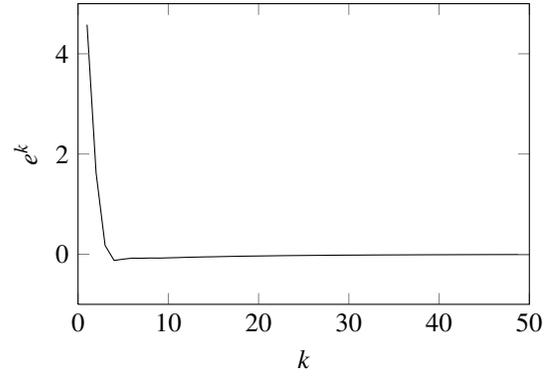
\begin{figure}[th!]
\centering
\pgfplotsset{width=ex2_error1_notRelZoom.tikz\columnwidth,height=ex2_error1_notRelZoom.tikz\columnwidth,compat=newest,plot coordinates/math parser=false}
%
%
\begin{tikzpicture}

\begin{axis}[%
scale only axis,
width=6cm,
height=4cm,
xmin=0, xmax=50,
ymin=-1, ymax=5,
xlabel={$k$},
ylabel={$e^k$},
axis on top]
\addplot [
color=black,
solid
]
coordinates{
 (1,4.58108)(2,1.62265)(3,0.174667)(4,-0.124377)(5,-0.0978099)(6,-0.078195)(7,-0.0802963)(8,-0.0753557)(9,-0.0767319)(10,-0.0717766)(11,-0.0674123)(12,-0.0610388)(13,-0.057115)(14,-0.0532295)(15,-0.0504687)(16,-0.0465072)(17,-0.0429136)(18,-0.039429)(19,-0.0371216)(20,-0.0348337)(21,-0.0325537)(22,-0.0299187)(23,-0.0277422)(24,-0.0259913)(25,-0.0246141)(26,-0.0230493)(27,-0.0213966)(28,-0.0198372)(29,-0.0186679)(30,-0.0176931)(31,-0.016703)(32,-0.0155769)(33,-0.0145206)(34,-0.0136711)(35,-0.0130071)(36,-0.0123345)(37,-0.0115862)(38,-0.0108416)(39,-0.0102342)(40,-0.009754)(41,-0.00929604)(42,-0.00878186)(43,-0.00825883)(44,-0.00781165)(45,-0.00745954)(46,-0.00713388)(47,-0.00677635)(48,-0.00640145)(49,-0.00606953)(50,-0.0058028) 
};

\end{axis}
\end{tikzpicture}
\caption{Error of cost function. The error for each iteration $k$ in ADMM is shown.}\label{fig:ex3_error}
\end{figure}

\begin{figure}[th!]
\centering
\pgfplotsset{width=ex3_residuals1vline.tikz\columnwidth,height=ex3_residuals1vline.tikz\columnwidth,compat=newest,plot coordinates/math parser=false}\input{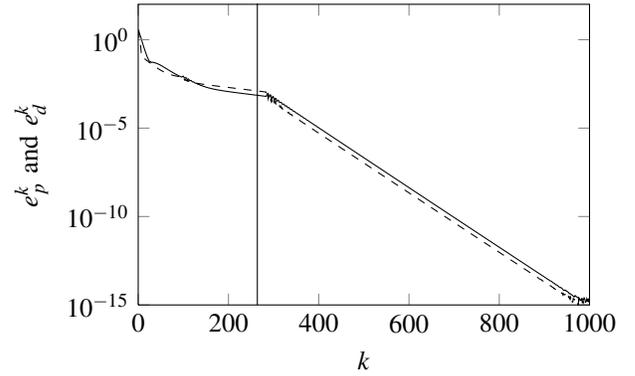}
\caption{Primal and dual residuals. The primal (\ref{solid}) and dual (\ref{dashed}) residuals are calculated for each iteration in ADMM in the first iteration of MPC. The vertical line shows where the stopping criteria are fulfilled. A rapid drop in the residuals occur in the first 20 iterations in ADMM.}\label{fig:ex3_residuals}
\end{figure}

Here we investigate the same set-up as in Example 1. We only consider $\lambda=0.1$ and the first optimization problem solved in the MPC iterations. We calculate the error of the cost function in \eqref{eq:mpc_cost} for each iteration in ADMM. The error is defined as $e^k = V^*-V^k$, where $V^*$ is the true optimal value of the cost and $V^k$ is the value obtained in ADMM iteration $k$. The true optimal value is approximated with the solution obtained by running ADMM for 1000 iterations. The optimal value is verified using CVX, a package for specifying and solving convex optimization problems, \cite{cvxProg}.  CVX calls the generic SDP solvers SeDuMi \cite{TTT:99} or SDPT3 \cite{Stu:99} to solve the problem. We choose to use SDPT3. The resulting error is shown in Figure \ref{fig:ex3_error}. The true optimal value is $V^*=4.58108$, and the final value obtained from ADMM is $V^{264}=4.58105$, where 264 is the number of iterations required to fulfill the stopping criteria. A rapid drop in the error occur in the first iterations in ADMM. The ADMM algorithm iterates until the stopping criteria are fulfilled, however, for improved visibility of the drop we only show the first 50 iterates. Note that since the ADMM solution is not necessarily feasible it is possible to achieve a value of the cost function at iteration $k$ that is lower than the optimal one. The corresponding primal and dual residuals are shown in Figure \ref{fig:ex3_residuals}. We see a rapid drop in error and residuals for the first 20 iterations in ADMM ($e^{20}=-0.03$, $e_p^{20}=0.11$ and $e_d^{20}=0.06$), confirming that ADMM converges fast to a moderate accuracy. 

\subsection{Example 4: Time of iterations in ADMM}
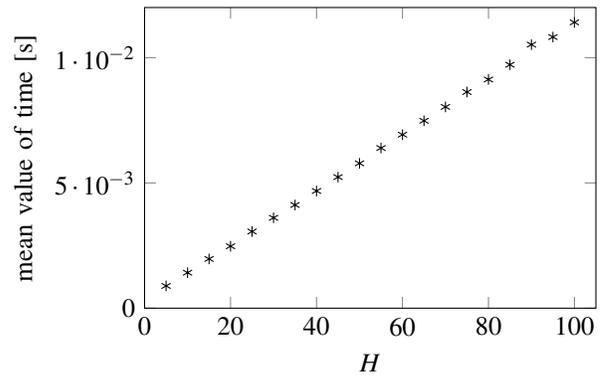
\begin{figure}[th!]
\centering
\pgfplotsset{width=ex5_mean100.tikz\columnwidth,height=ex5_mean100.tikz\columnwidth,compat=newest,plot coordinates/math parser=false}
%
%
\begin{tikzpicture}

\begin{axis}[%
scale only axis,
width=6cm,
height=4cm,
xmin=0, xmax=105,
ymin=0, ymax=0.012,
xlabel={$H$},
ylabel={mean value of time [s]},
axis on top]
\addplot [
color=black,
only marks,
mark=asterisk,
mark options={solid}
]
coordinates{
 (5,0.000887808)(10,0.00141636)(15,0.00196534)(20,0.00247168)(25,0.00305893)(30,0.00360703)(35,0.00411844)(40,0.00467786)(45,0.00523138)(50,0.00578156)(55,0.00639217)(60,0.00692554)(65,0.00748012)(70,0.00803397)(75,0.00862719)(80,0.00912829)(85,0.0097176)(90,0.01052)(95,0.0108214)(100,0.0114079) 
};

\end{axis}
\end{tikzpicture}
\caption{Mean value of time required for an iteration in ADMM. The mean values are calculated over 1000 iterations in ADMM, for a prediction horizon $H$ varying from 5 to 100 in steps of 5. The mean time is linear in the prediction horizon.}\label{fig:ex4_means}
\end{figure}

We consider the set-up in Example 1 with $\lambda=0.1$ and a prediction horizon $H$ varying from 5 to 100 in steps of 5. We only consider the first optimization problem solved in the MPC iterations and we fix the iterations in ADMM to 1000. We calculate the mean value of the time required for an iteration in ADMM. Figure \ref{fig:ex4_means} shows the resulting means with respect to the prediction horizon. We see that the mean time of the iterations in ADMM is linear in the prediction horizon. This is expected since the computational cost of the Riccati recursion is linear in $H$, \cite{rao1998}.

\subsection{Example 5: Required accuracy}
Example 3 shows how close the ADMM solution is to the optimal one for $\epsilon^{abs}=10^{-5}$ and $\epsilon^{rel}=10^{-4}$ in stopping criteria \eqref{eq:admm_stop}. However, the stopping criteria used may be to conservative with respect to required control performance. For example, if we in Example 1 with $\lambda=0.1$ restrict the number of iterations in ADMM to 10, we can have a sampling rate of 100 Hz in the MPC, see Figure \ref{fig:ex4_means}. The input signals obtained with both 10 and 1000 iterations in ADMM are shown in Figure \ref{fig:ex8_inputs}. The corresponding output signals are shown in Figure \ref{fig:ex8_outputs}. We see that, although the signals differ from each other, they still have the same over-all behavior. 
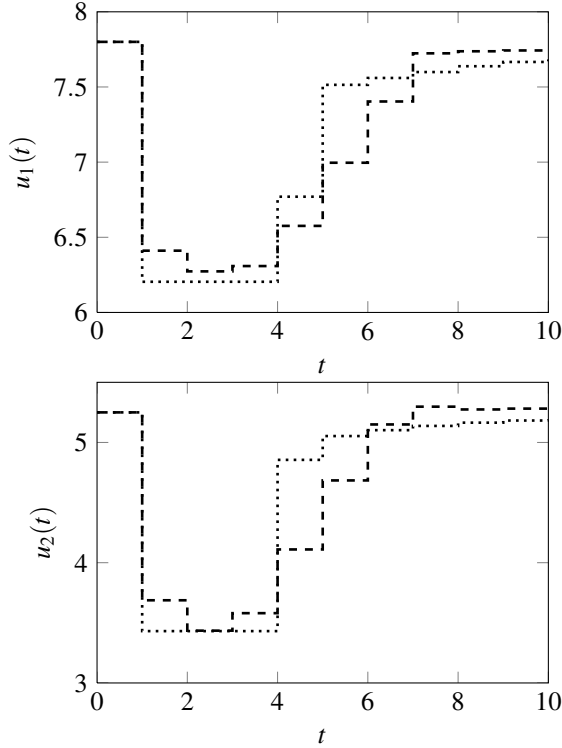
\begin{figure}[th!]
\centering
\pgfplotsset{width=ex8_input.tikz\columnwidth,height=ex8_input.tikz\columnwidth,compat=newest,plot coordinates/math parser=false}
%
%
\begin{tikzpicture}

\begin{axis}[%
name=plot1,
scale only axis,
width=6cm,
height=4cm,
xmin=0, xmax=10,
ymin=6, ymax=8,
xlabel={$t$},
ylabel={$u_1(t)$},
axis on top]
\addplot[const plot,color=black,dashed,line width=1.0pt] plot coordinates{ (0,7.8) (1,6.41098) (2,6.27328) (3,6.30861) (4,6.57582) (5,6.99597) (6,7.40341) (7,7.72374) (8,7.73741) (9,7.74289) (10,7.75016) };

\addplot[const plot,color=black,dotted,line width=1.0pt] plot coordinates{ (0,7.8) (1,6.20411) (2,6.20411) (3,6.20411) (4,6.77002) (5,7.51416) (6,7.55989) (7,7.59929) (8,7.63796) (9,7.66638) (10,7.6832) };

\end{axis}

\begin{axis}[%
at=(plot1.below south west), anchor=above north west,
scale only axis,
width=6cm,
height=4cm,
xmin=0, xmax=10,
ymin=3, ymax=5.5,
xlabel={$t$},
ylabel={$u_2(t)$},
axis on top]
\addplot[const plot,color=black,dashed,line width=1.0pt] plot coordinates{ (0,5.25) (1,3.68755) (2,3.43497) (3,3.58031) (4,4.11023) (5,4.6842) (6,5.14965) (7,5.29754) (8,5.27438) (9,5.28134) (10,5.28156) };

\addplot[const plot,color=black,dotted,line width=1.0pt] plot coordinates{ (0,5.25) (1,3.43112) (2,3.43112) (3,3.43112) (4,4.85511) (5,5.05312) (6,5.10211) (7,5.13743) (8,5.16453) (9,5.18227) (10,5.19131) };

\end{axis}
\end{tikzpicture}
\caption{Applied input sequences. The applied input sequence denoted (\ref{dashed}) and (\ref{dotted}) corresponds to 10 and 1000 iterations in ADMM respectively. The sequences differ, however the over-all behavior is the same.}\label{fig:ex8_inputs}
\end{figure}
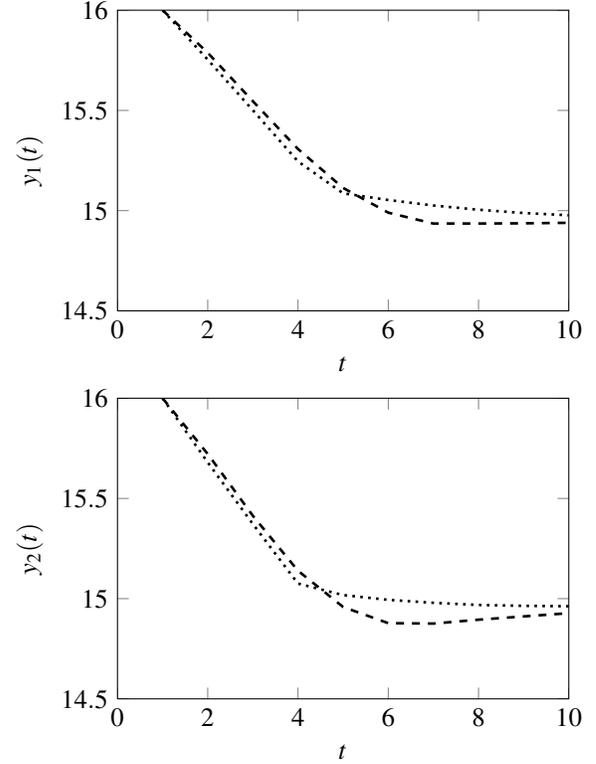
\begin{figure}[th!]
\centering
\pgfplotsset{width=ex8_output.tikz\columnwidth,height=ex8_output.tikz\columnwidth,compat=newest,plot coordinates/math parser=false}
%
%
\begin{tikzpicture}

\begin{axis}[%
name=plot1,
scale only axis,
width=6cm,
height=4cm,
xmin=0, xmax=10,
ymin=14.5, ymax=16,
xlabel={$t$},
ylabel={$y_1(t)$},
axis on top]
\addplot [
color=black,
dashed,
line width=1.0pt
]
coordinates{
 (1,16)(2,15.7878)(3,15.546)(4,15.3076)(5,15.1133)(6,14.99)(7,14.9355)(8,14.9357)(9,14.9367)(10,14.9393) 
};

\addplot [
color=black,
dotted,
line width=1.0pt
]
coordinates{
 (1,16)(2,15.7534)(3,15.5006)(4,15.2455)(5,15.0864)(6,15.0531)(7,15.0256)(8,15.0045)(9,14.9881)(10,14.977) 
};

\end{axis}

\begin{axis}[%
at=(plot1.below south west), anchor=above north west,
scale only axis,
width=6cm,
height=4cm,
xmin=0, xmax=10,
ymin=14.5, ymax=16,
xlabel={$t$},
ylabel={$y_2(t)$},
axis on top]
\addplot [
color=black,
dashed,
line width=1.0pt
]
coordinates{
 (1,16)(2,15.7217)(3,15.4135)(4,15.1393)(5,14.9585)(6,14.8777)(7,14.8756)(8,14.895)(9,14.9114)(10,14.9275) 
};

\addplot [
color=black,
dotted,
line width=1.0pt
]
coordinates{
 (1,16)(2,15.6798)(3,15.3724)(4,15.0752)(5,15.0179)(6,14.994)(7,14.9793)(8,14.9685)(9,14.9641)(10,14.962) 
};

\end{axis}
\end{tikzpicture}
\caption{Measured output sequences. The measured output sequence denoted (\ref{dashed}) and (\ref{dotted}) corresponds to 10 and 1000 iterations in ADMM respectively. The sequences differ, however the over-all behavior is the same.}\label{fig:ex8_outputs}
\end{figure}

\section{Conclusion}
We have derived a method for solving optimization problems with an $\ell_1$ regularized cost function subject to recursive equality constraints. The optimization problem occurs in control applications, e.g. $\ell_1$ regularized MPC. The method is based on the ADMM algorithm. We have showed that the costly projection step in ADMM is equivalent to solving an LQ regulator problem with an additional linear term in the cost function. Such problems can be efficiently solved using Riccati recursion. Future work consists of expanding the proposed method to $\ell_1$ regularized cost functions subject to both recursive equality and inequality constraints.

\appendix[Riccati recursion]
\label{sec:riccati}
We use a Riccati recursion to solve the projection problem \eqref{eq:admm_step2}, as in \cite{gla+jon84}. We showed in Section \ref{sec:proj} that the solution to \eqref{eq:admm_step2} is equivalent to the solution of a system of linear equations,
\begin{equation}\label{eq:riccati_kkt}
\left[
\begin{array}{ll}
\mathcal{Q} & \mathcal{A} ^T\\
\mathcal{A} &0 \\
\end{array}
\right]\left[
\begin{array}{l}
\xi\\
\lambda\\
\end{array}
\right]=\left[
\begin{array}{l}
r_{\xi}\\
r_{\lambda}\\
\end{array}
\right],
\end{equation}
where $\mathcal{Q}$ and $\mathcal{A}$ are block-diagonal matrices defined as
\begin{equation}\nonumber
\mathcal{Q}=
 \left[
\begin{array}{ll}
Q& 0\\
0 & \tilde{Q}\\
\end{array}
\right],\quad
Q=I_{H}\otimes
 \left[
\begin{array}{ll}
Q_{11}& Q_{12}\\
Q_{12}^T & Q_{22}\\
\end{array}
\right],
\end{equation}
and 
\begin{equation}\nonumber
\mathcal{A}=
 \left[
\begin{array}{llllll}
I & 0&0&0&\ldots&0 \\
-A & -B&I&0&\ldots&0 \\
\vdots & \ddots&\ddots&\ddots&\ddots&\vdots \\
0 & \ldots&\dots&-A&-B&I\\
\end{array}
\right].
\end{equation}
The vectors $\xi$, $\lambda$, $r_{\xi}$ and $r_{\lambda}$ can be divided into sub-vectors
\begin{equation}\nonumber
\begin{array}{ll}
\xi=(x_0,u_0,\ldots,u_{H-1},x_H),&\lambda=(\lambda_0,\ldots,\lambda_H),\\
 r_{\xi}=(r_{x,0},r_{u,0},\ldots,r_{u,H-1},r_{x,H}),&r_{\lambda}=(r_{\lambda,0},\ldots,r_{\lambda,H}),\\
\end{array}
\end{equation}
where $x_i$ is given by the system equations 
\[x_{i+1} = Ax_i+Bu_i+r_{\lambda,i+1}.\]
 It is shown in \cite{rao1998}, that there exists a matrix $S_i$ and a vector $\Psi_i$ such that 
\[
\lambda_i+S_ix_i=\Psi_i,\ i=0\ldots H,
\]
where $S_H=\tilde{Q}$, $\Psi_H=r_{x,H}$, and $x_0=r_{\lambda,0}$. The matrices $S_i$ and vectors $\Psi_i$ for $i=0,\ldots, H$ can be found through backward recursion. We then obtain $\xi$ and $\lambda$ through forward recursion. The algorithm is as follows \cite{rao1998}:\\
\emph{Backward recursion}: Update $S_i$ and $\Psi_i$,
\begin{equation}\nonumber
\begin{array}{lll}
&F_{i+1} = Q_{11} + A^TS_{i+1}A,&H_{i+1}= Q_{12} + A^TS_{i+1}B,\\
&G_{i+1} = Q_{22} + B^TS_{i+1}B,&\psi_{i+1}=\Psi_{i+1}-S_{i+1}r_{\lambda,i+1},
\end{array}
\end{equation}
\begin{align*}
&S_{i} = F_{i+1}-H_{i+1}G^{-1}_{i+1}H^T_{i+1},\\
&\Psi_{i}=r_{x,i}+A^T\psi_{i+1}-H_{i+1}G^{-1}_{i+1}(r_{u,i}+B^T\psi_{i+1}).
\end{align*}
\emph{Forward recursion}: Update $\lambda_{i}$, $u_{i+1}$ and $x_{i+1}$,
\begin{align*}
\lambda_i&=-S_ix_i+\Psi_i,\\
u_{i+1}&=G^{-1}_{i+1}(r_{u,i}+B^T\psi_{i+1}-H^T_{i+1}x_i),\\
x_{i+1} &= Ax_i+Bu_i+r_{\lambda,i+1}.
\end{align*}
\subsection{Unstable model}
If $A$ is unstable, the Riccati recursion might not provide the correct solution to \eqref{eq:riccati_kkt}. To avoid this, we pre-stabilize the state-space equations using state feedback control, see \cite{rao1998}, \cite{keerthi1986}, \cite{rossiter1998}. That is, we let
\begin{equation}\label{eq:feed}
\begin{array}{rl}
x_{i+1} &\!\!\!\!\!\!\! = Ax_i+Bu_i,\\
u_i&\!\!\!\!\!\!\!  = -Lx_i+v_k,
\end{array}
\end{equation}
where $L$ is the feedback vector. We can reformulate \eqref{eq:feed} as
\begin{align*}
x_{i+1} &= (A-BL)x_i+Bv_i,
\end{align*}
and treat $v_i$ as the unknown input signal. The solution obtained from the Riccati recursion will be the values of $x_i$ and $v_i$. The solution in terms of the original parameters $x_i$ and $u_i$, can be obtained as
\begin{align*}
\left[
\begin{array}{l}
x_i\\
u_i\\
\end{array}
\right]&=
\left[
\begin{array}{ll}
I&0\\
-L&I\\
\end{array}
\right]
\left[
\begin{array}{l}
x_i\\
v_i\\
\end{array}
\right].
\end{align*}

\bibliography{refs,myref}

\end{document}